# Critical Infrastructure in the Multi-Cloud Strategy: Use of Cloud Computing in SMEs

**Ruwan Nagahawatta**
**School of Accounting, Information Systems and Supply Chain**
RMIT University
Victoria, Australia.
Email: Ruwan.nagahawatta@rmit.edu.au

**Sachithra Lokuge**
School of Business
University of Southern Queensland
Queensland, Australia.
Email: ksplokuge@gmail.com

**Matthew Warren**
Centre for Cyber Security Research and Innovation
RMIT University
Victoria, Australia
University of Johannesburg, South Africa
Email: matthew.warren2@rmit.edu.au

**Scott Salzman**
Department of Information Systems and Business Analytics
Deakin University
Victoria, Australia.
Email: scott.salzman@deakin.edu.au

**Abstract**

Cloud computing enables cost-effective on-demand network access to a shared pool of configurable computing resources. The purpose of this paper is to examine and identifying the use of Cloud computing in the critical infrastructure domain among small and medium sized enterprises (SMEs). The data for this study were gathered from a survey of different academic, industry, governmental and online literature related to the use of Cloud computing in SMEs. The result revealed that there are risks involved in the use of Cloud computing, SMEs are deploying Cloud computing using different deployment models and reaching a high level of deployment within the critical infrastructure. The research findings are useful for SMEs that are planning or are in the use of Cloud computing, as well as for SMEs policymakers and business support community that engaged with Cloud computing initiatives.

# Introduction

Cloud technology is considered a service model for providing a variety of services called Cloud services. Nowadays, cloud-enabled solutions are widely used in various areas such as business, government, medical, education, and entertainment, etc. If enterprises adopt cloud services, they could save money, time and energy (Armbrust et al. 2010; Marston et al. 2011). Since these resources can be invested in other value-adding areas of their enterprises (Carroll et al. 2011), it offers businesses the ability to be productive and efficiency which is most important for companies such as small and medium sized enterprises (SMEs) (Carr 2005). Cloud computing brings effectiveness to the environment and economy since virtualization (Marston et al. 2011; Mudge 2010). Cloud service providers (CSPs) are experts in delivering computing services; consequently, they can achieve computing related tasks more effectively and economically (Armbrust et al. 2010). Furthermore, delivering any service on large scale brings in economies of scale which results in more effective operations. Economies comprising of more effective and state-of-the-art enterprises are certainly better off than those which are comprised of inefficient enterprises (Carroll et al. 2011). The small size firms should take advantage of these services, which will both enrich technology-enabled business, and importantly, reduce the cost impact on their business (Alshamaila et al. 2013; Nagahawatta et al. 2025).

Cloud computing helps to reduce the cost of IT infrastructure and services, improve productivity and efficiency and provide on-demand services. In addition, cloud computing adoption lets the organisations focus on their core businesses and let the IT burden be on the cloud computing providers and the nature of cloud computing provide the flexibility of starting small and grow or more precise resize as per the business needs. Thus, Cloud computing is still in its growing stages of diffusion; therefore studying its adoption process is very important. Indeed, cloud technology can deliver the most advanced software applications, hardware resources, and other services to both large and small size businesses. However, it seems that organisations are slow in accepting the cloud due to security issues and associated challenges (So 2011). Specifically, several studies have emphasized security and privacy related issues as major restraints for Cloud computing adoption. On the other hand, SMEs looking for the digital market place to reduce cost and improve ICT efficiency. In spite of the advantages, there is still resistance from enterprises to use of Cloud computing services.

Cloud technology can deliver the most advanced software applications, hardware resources, and other services to both large and small scale businesses. In reviewing literature have investigated a number of advantages that can be drive business to adopt cloud computing (Gupta et al. 2013). However, several studies have emphasized security and privacy issues as major resistant to cloud computing adoption. Further, scholars suggest security related elements and roles as integrated into cloud computing adoption and SMEs. Even though, security is the most critical factor to make a decision to use of cloud computing within SMEs. However, it is not a hesitation of using or not anymore but the question of how to mitigate the risks involved after the deployment.

# Critical Infrastructure

Critical infrastructure service includes water, wastewater, telecommunications, energy and transportation and other services (Brown et al., 2017). They are also stating that there is high interdependency between the critical infrastructure systems and vulnerable to cascading failures. The critical infrastructure systems are dynamic systems and reliant and influence each other and necessary to function together dynamically to supply the service normally in (Pye &Warren, 2007; Kaluarachchi et al 2020). The destruction to a single system has cascading effects on other systems within the critical infrastructure.

There are different sources for the destruction of the critical infrastructure. It can be destroyed, damaged or disrupted by breakdowns, negligence, accidents (Pye &Warren, 2007) natural disasters and extreme weather conditions, (Tsavdaroglou et al., 2018; Pye &Warren, 2007). In addition, the critical infrastructure can be impacted by human factors such as social engineering techniques (Ghafir et al., 2018). Since most of the critical infrastructure systems are based on information and communication technologies, cyber incidents are in relation to critical infrastructure can be a target for both conventional and information warfare (Cazorla et al., & Pye &Warren, 2007). The critical infrastructure is a vital asset for the maintenance of vital societal such as financial systems and power distribution networks (Merabti et al., 2011). In Cloud environment, critical infrastructure providers would require scalable platforms for their large amount of data and computation, multi-tenant billing and virtualization with very strong isolation, Service Level Agreement (SLA) definitions and automatic enforcement mechanisms, end-to-end performance and security mechanisms. However, these requirements might not be met by the CSPs as they suffer from some threats and challenges. Critical infrastructure sectors are migrating to Cloud computing to realise benefits such as scalability, high availability and decreased ownership cost (Office of Cyber and Infrastructure Analysis, 2017). Cloud computing can lead to faster delivery pace, continuous improvement cycles, broad services access, reduce maintenance effort and refocus that effort to improve service delivery (Digital Transformation Agency, 2017). SMEs in many countries are using Cloud computing in for their critical infrastructure.

## SMEs

The term SMEs is commonly used in many countries, world organisations and policymakers. Some other counties used the term small and medium businesses (SMBs) as well. The definition and classification of enterprises are commonly based on measurable factors such as the number of employees, the value of assets or sales volume (Rahman 2001). However, there is no generally accepted definition of small and medium scale enterprises and existing definitions vary from country to country, from sector to sector, over time, even among institutions within a single country. In other words, there is no specific definition for SMEs or SMBs in the world context. To define small and medium sized, industries generally use criteria such as the value of capital, amount of revenue, number of employees, estimate of fixed assets, and the nature of the business or combination of those. SMEs can be important in many ways. Small firms are vital to the global economy (Wiklund and Shepherd 2005). The most important and significant part of world economies is small businesses. Therefore, many policymakers and researchers are trying to understand these businesses. SMEs generally vary from large enterprises in terms of capacity, structure, and size of the business. The key challenge that SMEs face is keeping the costs under control since they are not able to spend a significant amount of money on their IT. Therefore, this contributes to the following common problems: lack of IT resources, lack of IT professionals and security practices (Welsh and White 1981). However, they have some advantages in terms of fast communication between managers and employees and their ability to execute and implement decisions quickly.

## Security issues of SME and Cloud

Cloud computing is a set of services concerned with accessing online applications, storage, platforms, and network through the Internet. Cloud Computing, which has been emerging over the past few years, is still in its growing stages and many challenges exist (Senarathna et al. 2018). Cloud technology can be realized as an innovative phenomenon which is a set of resolutions for the way the internet is used. There are several definitions for Cloud computing by institutes, experts and researchers. The simplest definitions for cloud computing given by using a key term to describe as computing services provided via the Internet (Katzan 2010). However, cloud computing does not have a global accepted standard definition, nevertheless, there are some generally identified definitions (Marston et al. 2011; Nagahawatta et al. 2019). One of the most common use definition for Cloud Computing is given by the national institute of standards and technology (NIST) as: "A model for enabling convenient, on-demand network access to a shared pool of configurable computing resources (e.g., network, servers, storage, applications and services) that can be rapidly provisioned and released with minimal management effort or service provider interaction" (Mell and Grance, 2011). Further, NIST provides a comprehensive conceptualization of cloud computing as five important characteristics, three cloud service models and four cloud deployment models to describe Cloud computing.

Cloud computing could offer the initial benefits for SMEs to attempt innovative software applications in a cost-effective manner (Carr 2005). Many SMEs are incapable to afford their own IT infrastructure but have an adequate IT budget according to their usage and requirements. Further, SMEs can reduce their capital expenditure for IT infrastructure by utilising and paying for the services and resources provided by the Cloud computing environment (Rittinghouse and Ransome 2009). Cloud computing offers appropriate business models and even adopted a combination of different business models to improve the performance of their businesses. In addition, Cloud computing can provide for SMEs' needs as direct and indirect go-to-market models. Moreover, Cloud Computing seems to be a more suitable option for many SMEs due to flexible cost structures and scalability (Sultan 2011).

SMEs can use Cloud computing for a variety of different applications such as corporate website, content management system (CMS), email, internal payroll processing, (customer-relationship management) CRM, archiving of internal corporate documents. Cloud computing can offer many business benefits for SMEs especially as Cloud services are most often "pay as you go", which can be an attractive cost structure for an SME, ICT experts and software, avoiding an upfront investment in hardware (Senarathna et al. 2018; Nagahawatta and Warren 2020). The overall cost when adoption is often less than the cost when going with traditional IT solutions (Dekker and Liveri 2015; Nagahawatta 2022). Online collaboration is often easier in the Cloud case as access is warranted to users from various end-user devices, physical locations, advantages for information security and network. Generally, speaking large CSPs can deliver advanced security measures, while spreading the connected costs across many customers.

The security attacks and threats are diverse in terms of motivation and technological exploits ranging from insider attacks motivated by malice to the accidental misconfiguration of enterprise networks, lack of contingency planning, to automated exploit of known security vulnerabilities. Cybersecurity strives to protect the confidentiality, availability and integrity of data and information from internal and external braches (Victorian Government, 2017). There are different possible risks in cloud services that are common with the traditional information technology model. The main difference between the two is that by default when a service is in the Cloud, it is accessible by any device connected to the internet. Examples of such threats are: brute force, data leakage, denial of service, domo escalation, hyper-jacking, phishing, RAM scraping ("A type of malware designed for monitoring and extracting data from a system during data processing while it is unencrypted") and virtual machine escape ("The act of escaping a virtual machine (a virtual system or application that is running inside a physical system) and interacting directly with the virtual machine's hosting environment.") (Office of Cyber and Infrastructure Analysis, 2017). Further, Cloud computing has carried its unique security concerns and threats such as Cloud malware injection attack, Metadata spoofing attack, Account and service hijacking, Unknown risk profile, Malicious insiders and Abuse and nefarious use of cloud computing (Younis & Kifayat, 2013; Nagahawatta et al. 2020). Also, there is the information risk which is the use of information power to the advantage of the attacker.

## Case Study: Compromise of an Australian small company with national security links

In November 2016, the Australian Cyber Security Centre (ACSC) understood that a small Australian company's network had been effectively attacked by a malicious cyber adversary via linking to national security projects. According to the ACSC analysis, the adversary had been in access to the network for a long period and taken a considerable percentage of data. As indicated by the analysis, the adversary was able to gain the access to the company's network misusing an internet-facing server, and then had moved across the entire network with the support of administrative credentials. This enabled them to install multiple web shells in the victims' network; a script that can be uploaded to a webserver empowering remote administration of the machine, to gain and maintain further access (ACSC Threat Report, 2017 p.50)

## Findings and Conclusion

Cloud computing is a service-based information technology model that is replacing the in-house asset-based information technology model and can be deployed in four different ways. That is public, private, hybrid and community cloud. Cloud services are three types of service which are infrastructure as service, platform as a service and software as a service. There are many advantages to Cloud computing which include reducing cost and improving efficiency.

SMEs usually do not have enough finances to spend so much on IT infrastructure, maintaining their hardware and upgrading software. Cloud Computing plays an important role to contribute SMEs to decrease their expenses and increase efficiency and performance. It enables SMEs to develop application-driven their needs with less price in the long-term. Although this technology has a significant benefit, Cloud Computing-related risks should be considered. Critical infrastructure providers are looking at others for facilitating and enjoying the cloud computing features. However, without appropriate solutions for considerable several security and privacy challenges, the use of Cloud computing will not succeed.

The use of Cloud computing in SMEs involves shifting critical infrastructures to the Cloud. The nature of cloud computing and the critical infrastructure make them vulnerable to cyber warfare activities and attacks but the risks involved in this shift are not stopping SMEs from using cloud computing in its operation and they are approaching the deployment differently. This study has reviewed significant issues related to Cloud computing security and analysed security requirements for various critical infrastructure providers. A reliable access control system is a crucial requirement to secure Clouds from unauthorised access. Access control systems in Cloud computing can be more complex and sophisticated due to heterogeneity, diversity of service and dynamic resources.

In conclusion, Cloud computing is been used by many SMEs and by highlighting the characteristics of the cloud computing and the critical infrastructure and possible risks involved in addition to common technology associated risks, this paper point out how SMEs are a potential cyberwarfare battlefield. Also, it highlighted that it is the question of how to mitigate the risks rather than to use or not. Thus, researchers are encouraged to research not only common risks but as well on different risk mitigation strategies to reduce the loss in case of such cyber warfare that targets Cloud computing services used by SMEs. This study identified important risks that can each SME deciding to use Cloud computing. Technology development risks and outsourcing opportunism risks can be described as key associated risks. In the future research can work on ways to mitigate security problems by reducing these risks, and proposing the best approach.